# Title

Policy consequences of the new neuroeconomic framework

# Authors

**A. David Redish*** (University of Minnesota)
Department of Neuroscience, University of Minnesota, Minneapolis MN 55455
ORCID ID: 0000-0003-3644-9072

**Henri Scott Chastain** (University of Minnesota)
Minnesota Center for Philosophy of Science, University of Minnesota, Minneapolis MN 55455
ORCID ID: 0009-0004-7239-8377

**Carlisle Ford Runge** (University of Minnesota)
Department of Applied Economics and Law School, University of Minnesota, St. Paul MN 55108
ORCID ID: 0000-0003-3739-6021

**Brian M. Sweis** (Icahn School of Medicine at Mount Sinai)
Department of Psychiatry, Department of Neuroscience, Icahn School of Medicine at Mount Sinai, New York, NY 10029
ORCID ID: 0000-0002-2123-5824

**Scott E. Allen** (Cornell University)
Department of Physics, Cornell University, Ithaca NY 14853
ORCID ID: 0000-0002-7399-3408

**Antara Haldar** (University of Cambridge)
Faculty of Law, Judge Business School, University of Cambridge, Trinity Lane, Cambridge CB2 1TN, United Kingdom

* Corresponding author

# Abstract

Current theories of decision making suggest that the neural circuits in mammalian brains (including humans) computationally combine representations of the past (memory), present (perception), and future (agentic goals) to take actions that achieve the needs of the agent. How information is represented within those neural circuits changes what computations are available to that system which changes how agents interact with their world to take those actions. We argue that the computational neuroscience of decision making provides a new microeconomic framework (neuroeconomics) that offers new opportunities to construct policies that interact with those decision-making systems to improve outcomes. After laying out the computational processes underlying decision making in mammalian brains, we present four applications of this logic with policy consequences: (1) precommitment to avoid falling into the trap of sunk costs, (2) media consequences for changes in housing prices after a disaster, (3) contingency management as a treatment for addiction, and (4) how social interactions underlie the success (and failure) of microfinance institutions.

# Keywords

Decision-making. Computational neuroscience. Contingency Management. Sunk Costs. Asset pricing. Microfinance.

# Paper

## Introduction: representation matters

The promise of neuroeconomics is that neural processes will be key to understanding economic behavior and thus will have policy implications (Camerer et al. 2005; Glimcher et al. 2008; Rangel et al. 2008; Redish 2022). Current neuroeconomic decision-making theories derive from computational neuroscience, which posits that neural circuits implement information processing (Marr 1982; Churchland and Sejnowski 1994; Dayan and Abbott 2001). Information about the past (memory) is combined with information about the present (perception) and teleological information (agentic goals) to take actions that achieve the needs of the agent (Rangel et al. 2008; Redish 2013; Mitchell 2023).

One of the most important discoveries in computational science over the last century is that *how one represents information changes what one can do with it* (Cormen et al. 1990). A library that organizes fiction books by author but non-fiction books by subject makes it easy to find all of the non-fiction books about chaos theory quickly, but hard to find all of the non-fiction books by James Gleick. Similarly, it is easy to find all of the books by Amy Tan quickly, but hard to find all of the books about mother-daughter relationships. Because people tend to research nonfiction topics but authors have a fan base, this sorting structure aligned to different scenarios is more efficient than a single generic sorting would be. Analogously, animals (including humans) have multiple decision systems which represent information differently, to be better aligned to different situations they might find themselves in (Dickinson and Balleine 1994; Niv et al. 2006; Gläscher et al. 2010; Kahneman 2011; Redish 2013).

All normative optimizations have process assumptions underneath them — the optimal actions to take are not independent of the processes that select them. The classical “rational model” of economics assumes that an agent follows a single-valued function over all possible outcomes, subject to an arbitrary budget constraint (Samuelson 1938; Debreu 1959; Mas-Colell et al. 1995). Implicit in this theory is the assumption that agents possess infinite resources of knowledge and abilities to find a global optimum. As pointed out by Simon (1972), sometimes it is better to take a pretty good action fast than to take the time to calculate the absolute best choice (say, if you are a zebra deciding which way to escape a lion bearing down on you). In this “bounded rationality” formulation, calculation time provides an additional cost to that utility function. Other models have been proposed that have memory or storage capacity limitations (Kahneman and Tversky 1979, 2000; Shleifer 2000; Akerlof 2005; Gigerenzer and Brighton 2009), but, taken to its full extent, this insight implies that representation matters — how data is stored and how information is transformed (through neural circuits) changes what the optimal actions are (Lieder and Griffiths 2019). In this chapter, we argue that the neuroscience of decision making provides a new framework within which to build economic models and theories. (See Levenstein et al. (2023) for a discussion of models, theories, and frameworks.)

Current theories of decision making suggest that mammalian brains (including human ones) consist of multiple neural circuits that represent and process information about the past, present, and future in different ways that are optimized for different world conditions. The argument put forward in this chapter is that the computational neuroscience of decision making provides a new microeconomic framework that offers (1) better descriptions of actual behavior, (2) new opportunities for policy that interact with those decision-making systems to improve outcomes, (3) better descriptions of the causes underlying the structures of group interactions, and (4) new understandings that can inform institutional policies because of their dependence on neural computations. In the following sections, we review the current neuropsychological theory of decision-making and then examine examples of these four consequences.

## A microeconomic framework derived from neuroscience

Agents are evolved to take actions in an uncertain but probabilistically predictable world (Mitchell 2023). As such, information (Shannon 1948) about the state of the world can be used to drive behavior; neural firing patterns that differ as a function of different conditions in the world allow effectors within the agent (such as muscles) to act (Rieke et al. 1997). For example, the Mauthner cell in fish detects stimuli that increase in size on the retina (looming stimuli), and the fish swims away from that stimulus (likely a diving predator!) very fast (Diamond 1971; Preuss et al. 2006). Importantly, the muscles in the fish are not responding to the predator, but to the firing of the Mauthner cell, which shares *mutual information* with the presence of a predator.

These neural circuits that transform information are computational processes with normative consequences. ***Memory*** in neural circuits, for example, arises through *content addressable* processes, such that successful retrieval is reactivation of a stored neural pattern (Hebb 1949; Hertz et al. 1991). Small changes can arise with each retrieval, which can be mediated by the cues that produced that recollection (Kahana 2020; Runge et al. 2023). These effects make memory unreliable (Loftus and Palmer 1974; Wells and Loftus 1984), but can also provide for behavioral flexibility (Mau et al. 2020; Sweis et al. 2021). ***Perception*** in neural circuits is fundamentally belief in a description of the world — we see trees and leaves, not green and brown pixels (Gibson 1977; Rao and Ballard 1999; Friston 2005; Adams et al. 2013; Sterzer et al. 2018). Perception entails a process of categorization and generalization through parallel distributed processing that depends on experience and expertise (Grossberg 1976; Hertz et al. 1991; McClelland and Rogers 2003; Pettine et al. 2023). ***Motivation*** and ***evaluation*** are computational memory processes of their own (Balleine and Dickinson 1991; Andermann and Lowell 2017; Sharpe et al. 2021). For example, if one is ill after eating food, one learns to dislike that food, which is why you should eat weird-flavored foods before getting chemotherapy (Bernstein and Webster 1980). Finally, the ***actions*** themselves depend on neural representations of the physics of the body ("forward models", (Kawato 1999)) and oscillatory processes that can step through those action sequences ("central pattern generators", (Grillner and Wallén 1985)).

Computationally, neural circuits affect the solutions available to the system. For example, cortical circuits implement *divisive normalization* that ensures maximal efficiency within the range of actual stimuli being experienced (Carandini and Heeger 2011). This is why we see white paper as white and black as black in both bright sunlight and a dim fluorescent-lit room, even though black paper in sunlight is actually thousands of times brighter than white paper in the fluorescent-lit room. Divisive normalization has been proposed to underlie range normalization in both normal economic behavior and after experience with extremes, such as after taking heroin (Webb et al. 2021; Gueguen et al. 2024). Similarly, because memory *pattern completion* processes depend on synaptic connections in distributed neural circuits that participate in multiple memories (Hebb 1949; Hopfield 1982; Hertz et al. 1991; Kahana 2020), storing new memories can interfere with older memories unless those older memories are retrieved and restrengthened (Rumelhart et al. 1986; McClelland et al. 1995; Kahana 2020). These retrieval/restrengthening processes can explain the effect of media on changes in housing prices seen after a disaster (Runge et al. 2023).

An important open question is the relationship between these computational processes and utility-based theories of value, given the complexities of the computation underlying the neural processes. In neuroscience, the single-valued utility function theory (Samuelson 1938; Von Neumann and Morgenstern 1944; Debreu 1959; Mas-Colell et al. 1995) manifests as the hypothesis of *common currency* — that there exists a unitary neural representation of value without maintaining representations of the situation, goods obtained, or the information process leading to the decision. Although some data support a central value representation (Padoa-Schioppa and Assad 2006, 2008; Kable and Glimcher 2009; Levy and Glimcher 2012), other data suggest incompatibilities with such a single-valued utility function (Berridge and Robinson 2003; Rangel et al. 2008; Redish et al. 2008; Kurth-Nelson and Redish 2010). In particular, how value is measured (willingness to pay, willingness to sell, as a choice between options) reliably produces preference orderings incompatible with a singular scale (Kahneman and Tversky 2000; Lichtenstein and Slovic 2006; Ahmed 2010).

Current theories of action-selection identify four fundamentally different information processes to take actions (Redish 2013). Although these go by different names in different neuropsychological literatures, most theories divide action-selection into these fundamental processes.

**Reflexes** are simple stimulus-response pairs in which both the stimulus-detection and responses are genetically wired into the agent. Neurophysiologically, reflexes are primarily functions of spinal cord (Sherrington 1906), in part because they evolved to respond quickly. Typical reflexes include things like throwing out your hands to catch yourself when you fall. Reflexes are usually ignored in economic models, but they can have important consequences on agentic control.

**Instinctual systems** (sometimes called ***Pavlovian actions***, or ***stimulus-stimulus associations***) entail a limited repertoire of actions that one can learn when to release (Pavlov 1927; Breland and Breland 1961; LeDoux and Daw 2018). In classical psychological theories, these systems transfer responses from an *unconditioned stimulus* to a *conditioned stimulus* (Domjan 1998). They involve subcortical neural circuits of amygdala, hypothalamus, nucleus accumbens shell, and periaqueductal gray (McNally et al. 2011; LeDoux and Daw 2018). They classically include fight or flight responses, but they also can be more complicated, such as mating dances. In humans, they include many prosocial behaviors (Redish 2022), such as threatening and submissive postures, laughing with friends, a preference for fairness, and parochial altruism, in which one is cooperative (altruistic) to one's kith and kin and wary (xenophobic) towards strangers. Economically, instinctual systems are often modeled as providing additional utility to simple outcomes (Dayan et al. 2006), which have been proposed to underlie sign tracking (whereby the conditioned stimulus takes on characteristics of the unconditioned stimulus) (Robinson and Flagel 2009; Colaizzi et al. 2020), a sensitivity to sunk costs (taking into account costs already spent) (Sweis et al. 2018a; Lind et al. 2023), and the endowment effect (excess value assigned to retaining what one owns) (Kahneman et al. 1991; Redish et al. 2016).

***Deliberative systems*** (sometimes called ***model-based***, ***action-outcome***, or ***planning systems***) entail a search through imagination and simulation of predicted consequences of one's actions, followed by evaluation, and then active action-selection (Niv et al. 2006; Buckner and Carroll 2007; van der Meer et al. 2012; Redish 2016). They involve prefrontal circuits setting goals, hippocampal circuits creating that imagined outcome, and orbitofrontal circuits evaluating those imagined possibilities (Stott and Redish 2014; Redish 2016; Rich and Wallis 2016; Tang et al. 2021; Knudsen and Wallis 2022). Current theories suggest a role for dorsal prefrontal circuits in selecting actions based on those evaluations (Cisek et al. 2009; Wallis and Kennerley 2011; Balewski et al. 2023). Because the imagined outcomes are processed serially (one at a time), deliberative systems are slow and variable in their execution (which options are examined in what order may change from encounter to encounter). It is important to recognize that deliberative systems are not economically rational in the classical sense as they depend on memory computations to determine what is searched and also on evaluation computations, which, as noted above, depend on experiential histories.

***Procedural systems*** (sometimes called ***model-free***, ***situation-action***, or ***habit systems***) entail a process of situation-recognition followed by release of a well-practiced action-chain (Rand et al. 1998; Jog et al. 1999; Dezfouli and Balleine 2012; Smith and Graybiel 2013; Mugan et al. 2024). The situation-recognition component depends on cortical sensory circuits, and the selection of the action-chain depends on dorsal striatal circuits. Some theories posit that procedural systems are economic in that they learn the value of taking an action in a situation and release those actions proportionally to that valuation (Samejima et al. 2005; Hikosaka et al. 2014), but these valuations are learned from past experiences, not derived from future expectations (Niv et al. 2006). Because of this, procedural systems are not sensitive to sudden changes in valuation and can drive perseverative motoric responses in over-learned conditions (Dickinson and Balleine 1994).

An important open question is how actions are selected when these decision systems are in conflict. While the actual computations that underlie this deconfliction are controversial (Daw et al. 2005; Kable and Glimcher 2009; McLaughlin et al. 2021), taking actions when the systems agree on the action to choose are easier and faster than when they are in conflict (Stroop 1935; Guitart-Masip et al. 2012; Redish 2016). It is also known that these systems interact in real behaviors. For example, deliberative tree-search is affected by Pavlovian evaluations that change under depression (Huys et al. 2012), and deliberative systems can train procedural systems over time (Klein 1999). Moreover, while many tasks can be accomplished by multiple systems, changes in environmental conditions can encourage the use of one system over another. This important observation provides an access point to how institutional policies can be used to affect decision-making.

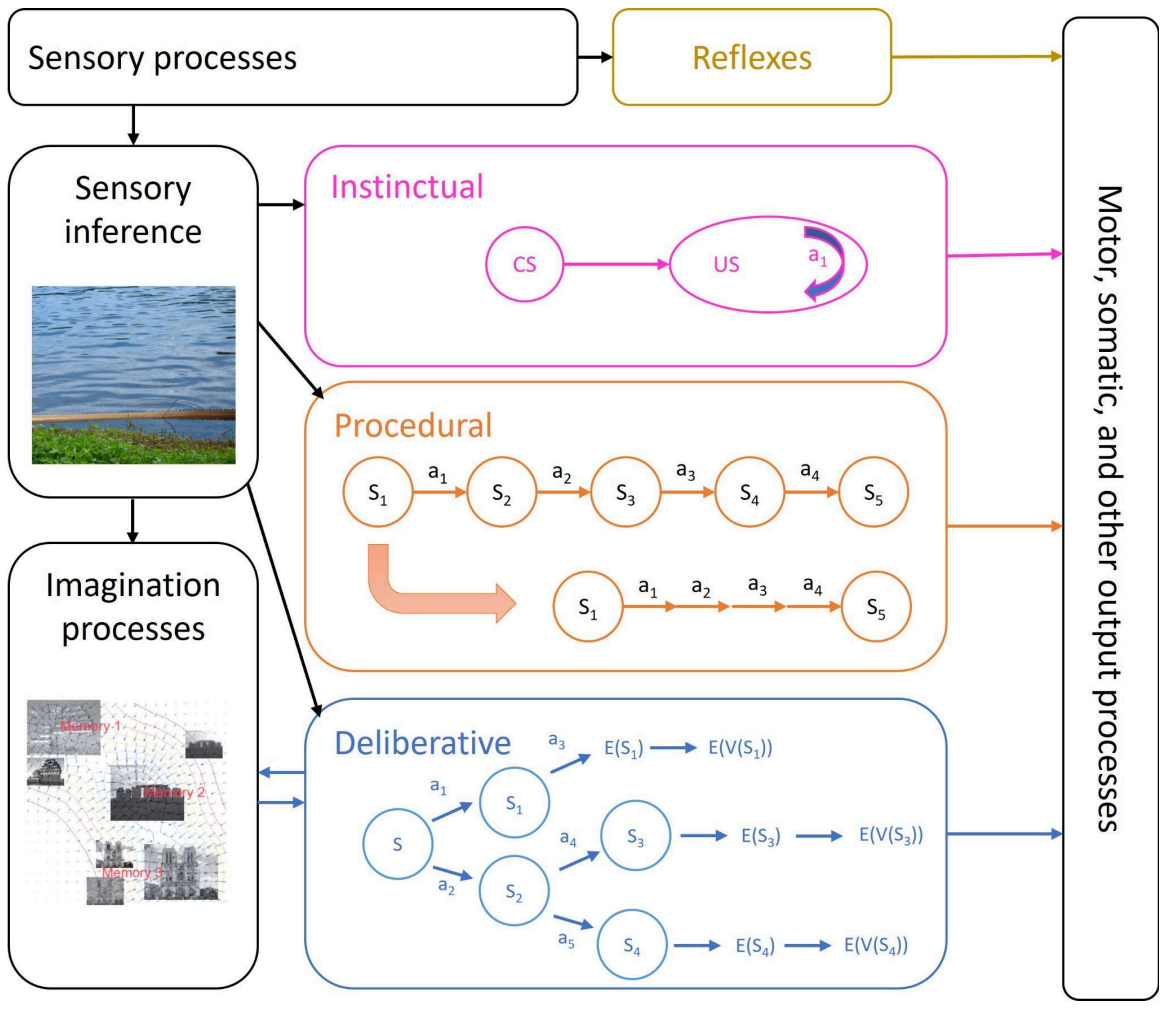

Figure: The four decision-making systems. Each system uses different representations and different learning processes applied to different information representations to select goals. The **instinctual** system associates conditioned stimuli (CS) with unconditioned stimuli (US) that releases an action *a*. The **procedural** system transforms a series of situation-action pairs into an action chain. The **deliberative** system works with memory and imagination to search through outcomes. These depend on **sensory interference** processes that see water and grass, not green and blue. **Imagination** works through content addressable processes. These all have consequences both for the development of preferences and for the expression of those preferences.

## Examples where the computational process matters

In the second half of this chapter, we explore descriptive and prescriptive consequences of this new neuroeconomic framework. First, we address the sensitivity to sunk costs in both human and non-human animals, showing that a computational explanation better explains the sunk cost fallacy and precommitment to avoid it than a sociological explanation does. Second, we show how the computational processes of memory can explain housing prices in response to floods and other disasters, and discuss how reminders can affect that memory process. While the observation that reminders improve memory is not novel, the computational process of memory explains how these reminders work. Third, we address how different scenarios impact preference ordering and suggest a novel explanation for contingency management as an addiction treatment, with novel policy consequences for improving contingency management. Finally, we discuss how computations underlying human prosociality affects the institution of microfinance, again with novel policy consequences. We have chosen these four examples to show the breadth and depth of this new neuroeconomic framework, from individual explanations for microeconomic observations (sunk costs) to policy systems that affect individuals (computational processes of memory, contingency management) to social policy and institutions (microfinance).

### Sunk costs and the endowment effect

Sunk costs are already spent and cannot be recovered. Because all futures include these same sunk costs, the optimal choice should be independent of those sunk costs. However, human choices often take sunk costs into account (Arkes and Blumer 1985; Mcafee et al. 2010; Sleesman et al. 2012). While many theories of the human sensitivity to sunk costs suggest a social cause (Staw 1976; Arkes and Blumer 1985; Kanodia et al. 1989; Staw and Ross 1989), recent work has found that non-human animals are also sensitive to sunk costs (Wikenheiser et al. 2013; Sweis et al. 2018a; Redish et al. 2022). In temporal serial foraging tasks, subjects choose to wait out a delay to a goal or to leave and proceed to the next opportunity. This willingness to quit depends on the time already spent waiting (the sunk costs) (Sweis et al. 2018a). In humans, that sensitivity to sunk costs depends on attention paid to the counting down delay (Kazinka et al. 2021). In mice, that sensitivity depends on the neural circuits involved in the Instinctual / Pavlovian action selection system, particularly the nucleus accumbens shell (Sweis et al. 2018b). Consistent with this hypothesis, in a different experiment, stimulation of the amygdala inputs to the nucleus accumbens shell increased sunk cost sensitivity in a conditioned place preference task (Lind et al. 2023).

*Conditioned place preference* is a behavioral phenomenon in which subjects prefer places where good things have happened to them over other places (Bardo et al. 1995; Tzschentke 1998; Mueller and de Wit 2011). Economically, conditioned place preference can be modeled as an increase in value of that place (Dayan et al. 2006), and may be analogous to the *endowment effect* seen in humans (Redish et al. 2016), in which people demand more money to sell an object than they are willing to pay for it (Kahneman et al. 1991). This hypothesis suggests that the endowment effect should depend on immediate, concrete experiences, because the instinctual system does not include a separate imagination component, and may be why concrete rewards are harder to forego (Bushong et al. 2010; Mischel 2014).

In temporal serial foraging tasks with a separate “offer zone”, where the delay does not count down, subjects do not include time spent in that offer zone in the sunk cost calculation (Sweis et al. 2018a). Mice, rats, and humans all learn to avoid wait-zone sunk costs, by making a separate decision in the offer zone (Sweis et al. 2018c, a; Kazinka et al. 2021; Huynh et al. 2021; Redish et al. 2022) This is a form of *precommitment*, in which one removes the opportunity to make a later decision that one thinks one will make “wrong” (Ainslie 1992). Precommitment requires multiple valuation functions; it cannot arise if the valuation function is unitary (Kurth-Nelson and Redish 2010). For example, one may recognize that if one goes to the party, one will drink (perhaps because of Pavlovian associations of alcohol and reward), thus one can deliberately decide not to go to the party. It is important to note that precommitment does not always entail deliberation overriding other systems. For example, learning the habit of driving home a different way to avoid the temptation of the casino is a process in which the procedural system provides the precommitment to avoid temptation (Elster 1999; Schüll 2012).

This neuroeconomic model not only provides explanations for the endowment effect and precommitment, it specifically identifies the mechanisms through which the decision systems interact to produce these effects and suggests policies (precommitment) that can change behavior.

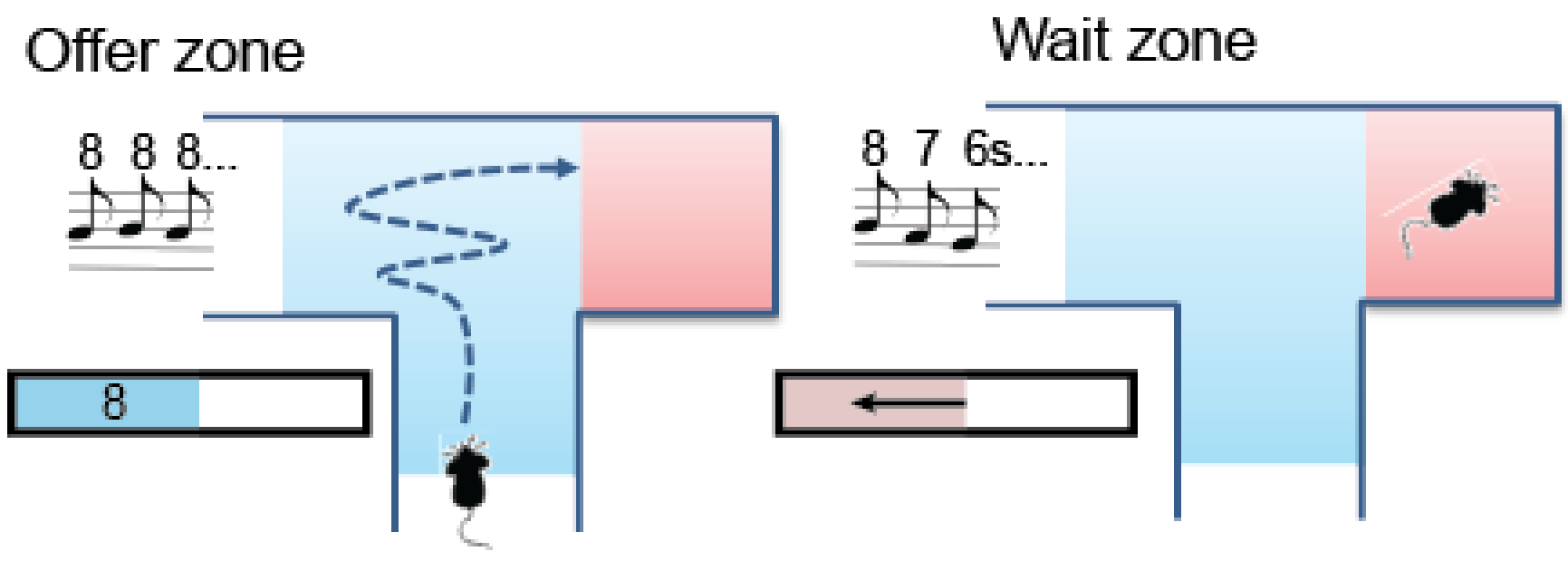

Figure: In the offer zone, subjects are informed of the delay, but the delay does not count down; in the wait zone, the delay counts down. In the offer zone, subjects dither and deliberate about whether to enter the wait zone or not. Subjects show a sensitivity to sunk costs in the wait zone, but not the offer zone. Rodents hear a tone to inform them of the delay; humans get a download waitbar to inform them of the delay. All subjects learn to precommit in the offer zone to avoid the sunk cost fallacy in the wait zone. See (Sweis et al. 2018c, a; Kazinka et al. 2021; Huynh et al. 2021; Redish et al. 2022).

## Memory implications

The computational processes that underlie perception and memory also have economic consequences. Current theories suggest that memory depends on *pattern completion* in which Hebbian learning strengthening synapses that connect co-active neurons produces dynamic changes in neural firing such that a partial (or nearby) pattern will return to an originally stored pattern (Hebb 1949; Hopfield 1982; Hertz et al. 1991; Kahana 2020). This recall process has consequences.

Both neurons and synapses participate in multiple memories, which implies that newly stored memories can *interfere* with older memories, moving the synapses away from the strengths needed to recall those older memories (Hasselmo 1993; McClelland et al. 1995). When combined with a temporal context model that contains slowly changing representations of time (Howard et al. 2005), new experiences interfere with the recall of old memories. Runge et al. (2023) built a computational model of single event storage (such as a flood) and learned valuation (remember that valuation is its own memory process) and found that both assessed and remembered cost decayed with a set time course. Remembered cost decayed to a minimum level, while assessed cost eventually decayed down to baseline. Importantly, however, reminders could disrupt this decay, producing long-term memories of the disaster that did not fade away.

This memory model of disaster suggests that media reminders may be critical to this time course. Gallagher (2014) found that flood-induced declines in housing prices depended more on whether a household was in a given media market than on whether the house itself was damaged by the flood. Reminders of disasters (such as recognizing the anniversary of a disaster) keep the memory alive, while forgotten dangers (and a lack of reminders) make a mitigated disaster ignored (such as the Y2K bug, which was only not a disaster because of the expensive work done to prevent it, and has now been mostly forgotten) (Runge et al. 2023). Furthermore, those reminders can change the narrative, and even the memory itself (Loftus and Banaji 1989; Sacchi et al. 2007).

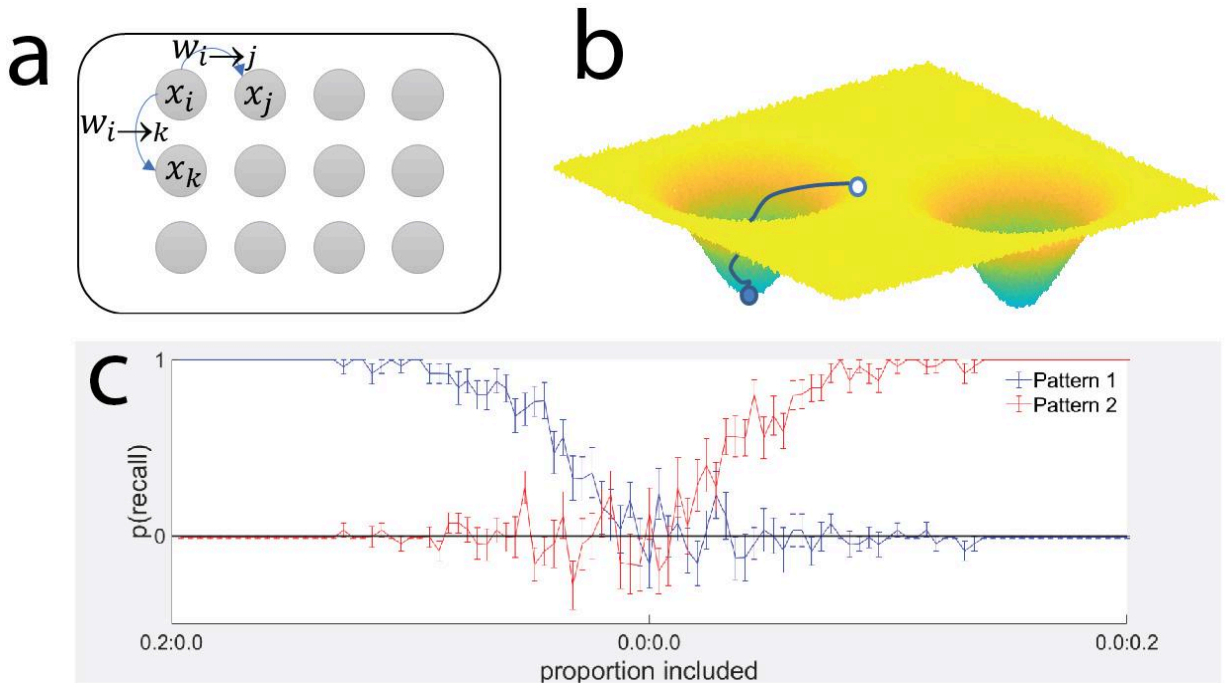

Figure: Memory processes. (a) Current theories of human memory are based on attractor networks of parallel distributed processes, with the memories stored through the changing of synaptic weights between units. (b) Mathematically, memories can be understood as “valleys” in a memory space. A nearby representational pattern is transformed through those weights to find itself in the valley. This means that (c) the likelihood of falling into a memory is dependent on how deep the valley is (proportional to the strength of the weights) and can be reinstated (and strengthened) with reminders, such as from news media. Modified after (Runge et al. 2023)

## Conflicting decisions (addiction and addiction treatment)

Addiction has been described as an overvaluation of drugs (Stigler and Becker 1977; Becker and Murphy 1988; Robinson and Berridge 2003; Redish 2004; Heyman 2009). However, measuring value in different ways can produce changes in preference orderings: measuring *willingness to pay* can reveal a different preference order than asking subjects to *choose between options* (Lichtenstein and Slovic 2006). For example, rats will press a lever much more for cocaine or heroin than they will for saccharin. However, when given an option to choose between drug or saccharine, many of those same rats choose saccharine over drug (Ahmed et al. 2002; Lenoir and Ahmed 2007; Lenoir et al. 2007; Cantin et al. 2010). Similar results can be found for social rewards vs drugs of abuse (Martin and Iceberg 2015; Venniro et al. 2018; Chow et al. 2022).

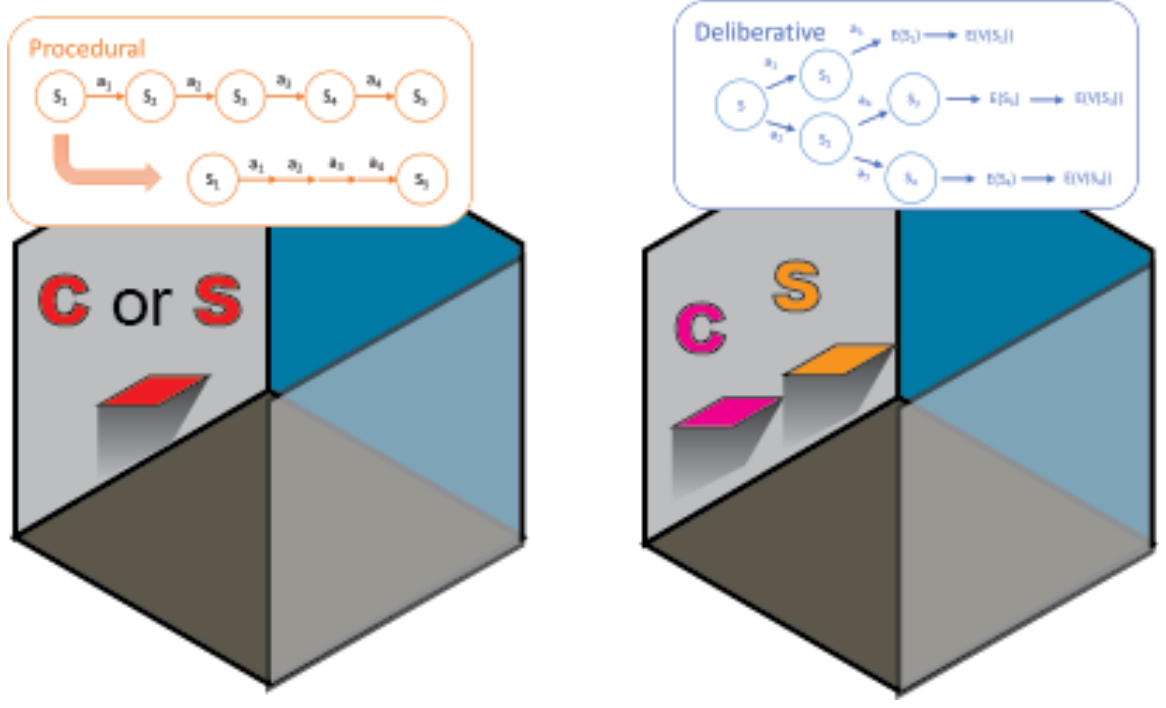

Figure: When animals only have one option, and the experiment measures the number of lever presses the animal will make to receive a reward (cocaine or saccharine), this is measuring the *willingness to pay* and finds that the WTP(C) > WTP(S), allowing us to infer that C > S. In contrast, when animals have two options, and the experiment measures which option is chosen, this is measuring preferences through *choose between* processes and finds that S > C. These differences arise because the willingness-to-pay activates the procedural system, while choose between activates the deliberative system.

The neuroeconomic explanation is that different behavioral decision-systems drive behavior in each scenario (Redish et al. 2016). The single-lever progressive-ratio experiment is most likely driven by procedural decision systems. (It is a single repeated action in response to a specific situation that escalates with experience. It is insensitive to outcome devaluation.) However, procedural decision systems are not well-suited to choosing between separate options. The two-lever choose-between scenario is much more likely driven by deliberative decision systems. Because these are two separate neural systems with separate evaluation processes, they can reveal incompatible preference orderings.

We can use this change in preference ordering to treat addiction.

*Contingency management* is one of the most successful treatments available for addiction (Higgins et al. 1991; Lussier et al. 2006; Petry 2011). In *contingency management*, a subject is provided a small (often monetary) reward for not using drugs in a recent time frame (typically a week). The original explanation for contingency management's success was economic — losing the reward was an opportunity cost that made the drug more expensive (Hursh 1991; Higgins et al. 1991). However, contingency management works better than this economic hypothesis would predict (Regier and Redish 2015; Davidson et al. 2024). An alternative explanation is that contingency management shifts the decision systems that the subject is using from either instinctual (drawn towards the drug) or procedural (willing to pay) systems to deliberative systems (choose between), much like the rats in the different experiments.

This proposed explanation for contingency management has important implications for implementation — variations that improve deliberative decision making should improve contingency management. Deliberation depends on imagined expectations of outcome and concrete outcomes are easier to imagine (Peters and Büchel 2010), so more concrete outcomes should be more effective. They are (Petry et al. 2015). Deliberation depends on imagining outcomes, so improving one's ability to imagine future outcomes (through *episodic future thinking training*) should reduce drug use and relapse. It does (Snider et al. 2016; Stein et al. 2016). Deliberation depends on prefrontal cortical circuits interacting with hippocampus, accumbens core, and other structures. Improving those connections should reduce drug use and relapse. It does (Camchong et al. 2023).

Importantly, it is not that deliberative decisions are always better, but rather that, for those subjects whose addictions are driven by their procedural or instinctual decision systems, contingency management can help them make deliberative decisions in a compensatory manner. Other subjects whose addictions are driven by deliberative dysfunctions will likely need other treatments, perhaps by encouraging new habits (procedural processes) (Redish et al. 2008).

This new process-based model provides us not only a new explanation of preference changes, but also suggests policies that can have direct effects on behavior.

## Microfinance

Humans are social animals and have evolved systems and developed technologies that increase prosociality (Wilson 2002, 2015; Fehr and Fischbacher 2003; Bowles and Gintis 2013; McCullough 2020; Redish 2022; Cushman 2024). To achieve prosociality, people make themselves vulnerable to others, whether through risking loss or foregoing gain. These actions depend on how the decision systems interact with the external social structure (Allen et al. 2024). Instinctual trust will be based on kith and kin, on cooperation within one's community, but will include a wariness of strangers. Instinctual trust is unlikely to keep track of explicit transactions, but likely depends on regular social gatherings to create community. Deliberative trust depends on direct expectations of imagined outcomes and thus is often explicitly transactional. Procedural trust depends on the practiced regularity of behavior (as might occur in a sports team or a surgical team). Institutions that access all three systems will be more likely to foster prosocial behaviors.

As a concrete example, we can look at microfinance, which succeeded in providing loans to communities not normally reached by the formal banking sector (particularly in South Asia). We can compare two specific microfinance institutions: Grameen Bank and SKS India. Both institutions provided similar services to similar communities with similar historical and ethnic backgrounds, structured in similar ways (Haldar and Stiglitz 2016). However, Grameen Bank has a strong track record of repayment (90% repayment over more than 25 million loans), while SKS India collapsed (facing default rates of around 90%). Grameen Bank leveraged "group lending", providing microloans ($500-$1000) to poor women (often in rural areas) organized into groups of 5-10 that met weekly with public repayment of loans (Bornstein 1999; Yunus 2007; Haldar and Stiglitz 2013). The economic literature interpreted the microfinance success in terms of its joint liability structure (Stiglitz 1990) — in theory at least, the Grameen Bank in Bangladesh provided loans to only some of the women in a group and would not provide additional loans until the first were repaid. When SKS Microfinance in India tried to replicate the joint liability model, however, it led to mass defaults.

The differences between these two banks suggests that the microfinance depended more on social pressure stemming from active peer-monitoring between borrowers and a robust relational contract between the bank and borrower than on the rational calculations of the joint liability. These strong community bonds incentivized the repayment of loans in the Grameen Bank but not SKS India. Compared to conventional formal banks, Grameen Bank deals were "informal" (requiring neither the provision of collateral by borrowers, nor formal contracts), which likely led to a more instinctual trust structure rather than transactional deliberative trust. Non-transactional deliberative trust was strengthened by the resilience of Grameen Bank in the face of the natural disasters of 1998. While SKS India ostensibly maintained this informal structure, it switched from a participatory model focused on poverty-reduction in the community to a for-profit model that made the lending relationship transactional rather than prosocial. Grameen Bank actively engaged instinctual trust with meaningful ritual and building of group social identity (weekly meetings, public repayment, reputation mechanisms, celebration of repayment within each group and socially punishing bad actors, and active peer monitoring that lowered information costs), until repayment became a matter of procedural trust. SKS India, on the other hand, relied on deliberative rather than instinctual trust, failing to create the community ties that activate the latter.

Experimental paradigms have found that when playing with the same people over and over again (i.e. living in a community), groups bifurcate, as more cooperators are identified, people become more willing to cooperate, while as more cheaters are identified, people become less willing to cooperate (Marwell and Ames 1981; Fehr and Fischbacher 2003). So observing that the others in your community are repaying the loans makes repaying them both the social norm of the community (instinctual), the thing one does (procedural), and the expectation that would be socially punished if not done (deliberative).

These neuroeconomic principles have direct consequences for how institutional policies change our behavior. Institutions that align with neuroeconomic principles that drive cooperation will drive people to cooperate, while misaligned institutions can have devastatingly negative consequences because they have made the wrong microeconomic assumptions. Importantly, while group lending was responsible for cooperative outcomes in the Grameen Bank, cooperation is not guaranteed by mechanically grouping individuals together; instead, groups need sufficient social salience to impact decisions: these nuances point to a need for new microeconomic models able to accommodate this socio- and neuroeconomic complexity.

## Conclusions, consequences, and predictions

All normative optimizations contain process assumptions. Conversely, how data is stored and processed changes what the optimal choices will be. Neuroeconomics says that the computational processes implemented by neural systems that produce behavior provide us a new framework with which to understand microeconomics, and suggest new models dependent on those computational processes.

In particular, the key prediction of this new framework is that the effects of behavioral manipulations will depend on which decision processes they access and how the behavioral manipulations interact with the underlying computational processes. Thus, for example, whether moral decisions are deontological or utilitarian in the classic trolley problem are driven by which decision process is activated (Greene et al. 2001; Moll et al. 2001); different framings are more likely to activate one decision process over the other, but it is the decision process that explains the behavior, not the framing (Costa et al. 2014). Some decision processes are sensitive to sunk costs, while others are not (Sweis et al. 2018b, a; Lind et al. 2023). A similar prediction can be made about loss aversion, the endowment effect, and other behavioral phenomena.

Furthermore, as we saw in the asset memory and contingency management examples, the specific computational processes underlying these behaviors have consequences as well. For example, making the second option more concrete makes it easier to imagine (Trope and Liberman 2003), more likely to be pulled up in a search process (Kurth-Nelson et al. 2012), and thus slows down discounting rates (Peters and Büchel 2010). Training someone to imagine future outcomes more deliberately can also slow down discounting rates (Radu et al. 2011) with consequences on drug demand curves (Snider et al. 2016; Stein et al. 2016). As another example, deliberative processes are based on sampling algorithms (Redish 2016), and thus, like all sampling processes, will be inconsistent, depending on which choices are sampled in which order (Stewart et al. 2006; Redish and Johnson 2007); in contrast, procedural processes are based on well-practiced expertise and produce reliable, consistent action chains (Rand et al. 1998). This means that framing that pushes someone to be more deliberate in their choices (listing pros and cons) will likely produce more variable preferences than well-practiced procedural processes.

These consequences do not diminish with larger scale policy questions, and thus, if one is going to attempt to create working institutions (such as microfinance lending institutions), one needs to take these computational processes into account.  These computational processes have consequences both for our understanding of economic behavior and for the institutional policies that affect it. The neuroeconomic framework provides the entry point into these computational processes.